\documentstyle[12pt]{article}
\begin{document}
\vskip 0.1in
\centerline{\Large\bf Anisotropic Born-Infeld Cosmologies} 
\vskip .7in
\centerline{Dan N. Vollick}
\centerline{Department of Physics}
\centerline{Okanagan University College}
\centerline{3333 College Way}
\centerline{Kelowna, B.C.}
\centerline{V1V 1V7}
\vskip .9in
\centerline{\bf\large Abstract}
\vskip 0.5in
Anisotropic cosmological spacetimes are constructed from spherically
symmetric solutions to Einstein's equations coupled to nonlinear
electrodynamics and a positive cosmological constant. This is accomplished by 
finding solutions in which the roles of $r$ and $t$ are interchanged 
for all $r>0$
(i.e. $r$
becomes timelike and $t$ becomes spacelike). Constant time hypersurfaces have 
topology $R\times S^2$ and in all the spacetimes considered the radius of
the two sphere vanishes as $t$ goes to zero. The scale factor of the 
other dimension diverges as $t$ goes to zero 
in some solutions and
vanishes (or goes to a constant) in other solutions. At late times local
observers would see the universe to be homogeneous and isotropic.
\newpage
\section*{Introduction}
Over the last few years Born-Infeld theory \cite{Bo1} has undergone a 
revival due to its appearance in string theory \cite{Po1}. In this paper
some exact cosmological solutions are found to the Einstein field equations
coupled to nonlinear electrodynamics and a positive cosmological constant.
These solutions are constructed from spherically symmetric solutions with
$g_{tt}=1/g_{rr}=-(1-2m(r)/r)$. If $m(r)>\frac{1}{2}r$ for $0<r<\infty$ then
$r$ and $t$ interchange roles and the solutions describe cosmological
spacetimes with a singularity at $t=0$ (instead of at $r=0$).
Constant time hypersurfaces have topology $R\times S^2$ and the radius
of the two sphere goes to zero as $t$ goes to zero. The scale factor of the
other dimension diverges as $t$ goes to zero in some solutions and vanishes
(or goes to a constant) in other solutions. The Schwarzschild solution
with a cosmological constant leads to a cosmological solution as does 
Born-Infeld theory. However, Maxwell's theory does not as it is not
possible to satisfy $m(r)>\frac{1}{2}r$ for all $r$ on $(0,\infty)$. Some other
Born-infeld cosmologies can be found in \cite{Gi1,Ga1}.
\section*{Born-Infeld Theory}
In nonlinear electrodynamics the Maxwell Lagrangian
\begin{equation}
L=-\frac{1}{4}F^{\mu\nu}F_{\mu\nu}=\frac{1}{2}(E^2-B^2)
\label{Maxwell}
\end{equation}
is replaced by
\begin{equation}
L=L(F^2,G^2)
\end{equation}
where $F^2=\frac{1}{2}F^{\mu\nu}F_{\mu\nu}$, $G^2=\frac{1}{2}F^{\mu\nu}F^{*}_{\mu\nu}$,
$F_{\mu\nu}^{*}$ is the dual of $F_{\mu\nu}$,
and $L$ is any function that reduces to (\ref{Maxwell}) in the weak
field limit. Born and Infeld took $L$ to be given by
\begin{equation}
L=-\frac{1}{a^2}\left[\sqrt{1+a^2F^2}-1 \right]
\end{equation}
For the solutions considered in this paper $\vec{B}=0$ so that $G^2=0$.
Thus, all $G^2$ dependences will be dropped.
  
The field equations are
\begin{equation}
\nabla_{\mu}P^{\mu\nu}=0
\end{equation}
and
\begin{equation}
\nabla_{\mu} F^{*\mu\nu}=0 ,
\end{equation}
where
\begin{equation}
P^{\mu\nu}=\frac{\partial L}{\partial F_{\mu\nu}}.
\end{equation}
The energy-momentum tensor is
\begin{equation}
T^{\mu\nu}=-2P^{\mu\alpha}F^{\nu}_{\;\;\alpha}+g^{\mu\nu}L
\end{equation}
and the ``Hamiltonian", which is a function of $P^{\mu\nu}$, is
\begin{equation}
H=P^{\mu\nu}F_{\mu\nu}-L .
\end{equation}
For the Born-Infeld Lagrangian
\begin{equation}
T^{\mu\nu}=\left[\frac{F^{\mu\alpha}F^{\nu}_{\;\;\alpha}}
{\sqrt{1+a^2F^2}}-\frac{1}{a^2}g^{\mu\nu}\left(\sqrt{1+a^2F^2}-1\right)
\right]
\end{equation}
and
\begin{equation}
H(P^2)=\frac{1}{a^2}\left[\sqrt{1+a^2P^2}-1\right]
\end{equation}
where $P^2=-2P^{\alpha\beta}P_{\alpha\beta}$.
\section*{Cosmologies from Spherically Symmetric Solutions}
Birkhoff's theorem holds for nonlinear electrodynamic theories and the
general spherically symmetric solution is \cite{Ho1,Ho2,Ho3,Pe1,De1,Wi1,Ol1}
\begin{equation}
ds^2=-\left[1-\frac{2m(r)}{r}\right]dt^2+\left[ 1-\frac{2m(r)}{r}\right]^{-1}
dr^2+r^2d\Omega^2
\end{equation}
\begin{equation}
P=\frac{Q}{r^2}dt\wedge dr
\end{equation}
and 
\begin{equation}
\frac{d m(r)}{dr}=4\pi r^2H(P^2)+\frac{1}{2}r^2\Lambda
\end{equation}
where $P^2=Q^2/r^4$ and $\Lambda$ is the cosmological constant.
  
If $m(r)>\frac{1}{2}r$ for $0<r<\infty$ then $r$ is a timelike coordinate
and $t$ is a spacelike coordinate. Relabeling $r$ and $t$ and denoting
the spacelike variable by $x$ gives
\begin{equation}
ds^2=-\left[ \frac{2m(t)}{t}-1\right]^{-1}dt^2+\left[\frac{2m(t)}{t}-1\right]
dx^2+t^2d\Omega^2
\end{equation}
\begin{equation}
P=\frac{Q}{t^2}dx\wedge dt
\end{equation}
and
\begin{equation}
\frac{dm(t)}{dt}=4\pi t^2H\left[\frac{Q^2}{t^4}\right]+\frac{1}{2}t^2\Lambda .
\label{mdot}
\end{equation}
Constant timelike surfaces have topology $R\times S^2$ and the two sphere
has radius $t$.
The Ricci scalar is given by
\begin{equation}
R=-2\left[\frac{t\ddot{m}+2\dot{m}}{t^2}\right] 
\end{equation}
and $R$ generically diverges as $t$ goes to zero.
  
Equation (\ref{mdot}) can be written as
\begin{equation}
\frac{dm(t)}{dt}=4\pi t^2H\left[\frac{Q^2}{t^4}\right]+\frac{1}{2}t^2\Lambda
\end{equation}
Integrating gives 
\begin{equation}
\frac{2m(t)}{t}-1=\frac{8\pi}{t}\int t^2H\left[\frac{Q^2}{t^4}\right]dt
+\frac{2m_0}{t}+\frac{1}{3}\Lambda t^2-1
\label{3}
\end{equation}
where $m_0$ is a constant. It is important to remember that the constraint
\begin{equation}
\frac{2m(t)}{t}-1>0
\end{equation}
must be satisfied for all $t>0$.
  
First consider the case $Q=0$ and take $H(0)=0$. The constraint becomes
\begin{equation}
\frac{2m_0}{t}+\frac{1}{3}\Lambda t^2-1 > 0.
\end{equation}
This will be satisfied if $\Lambda >0$ and $m_0 >\frac{1}{3}\Lambda^{-1/2}$. 
Even though $R$ remains finite as $t$ goes to zero the scalar
$R_{\mu\nu\alpha\beta}R^{\mu\nu\alpha\beta}$ diverges, so that $t=0$
is an initial singularity.
Thus, Schwarzschild with a positive cosmological constant can be converted into a
cosmological solution with metric
\begin{equation}
ds^2=-\left[\frac{2m_0}{t}+\frac{1}{3}\Lambda t^2-1\right]^{-1}dt^2+
\left[\frac{2m_0}{t}+\frac{1}{3}\Lambda t^2-1\right] dx^2+t^2d\Omega^2.
\end{equation}
As $t\rightarrow 0$ the two sphere collapses but the x direction
blows up. For large $t$ the metric is
\begin{equation}
ds^2=-d\tau^2+\exp\left[2\sqrt{\frac{\Lambda}{3}}\tau\right]\left[d\bar{x}^2
+d\Omega^2\right] ,
\end{equation}
where $\tau=\sqrt{3/\Lambda}\ln t$ and $\bar{x}=\sqrt{\Lambda/3}x$. Thus,
at late times we have inflationary behaviour and the scale factor of
the two sphere is the same as the scale factor for the x direction.
  
Next consider Maxwell's theory with $H(P^2)=1/2P^2=Q^2/2t^4$. The constraint is
\begin{equation}
\frac{2m_0}{t}-\frac{4\pi Q^2}{t^2}+\frac{1}{3}\Lambda t^2-1 > 0
\end{equation}
which cannot be satisfied for all $t>0$. The problem is that the $Q^2$ term
diverges faster than the $m_0$ term and has the wrong sign. This can be
modified in nonlinear electrodynamics by including a more divergent term
with the correct sign or by eliminating the divergence.
If Maxwell's theory is modified so that $H(P^2)=\frac{1}{2}P^2-\alpha^2 P^4$,
the constraint becomes
\begin{equation}
\frac{2m_0}{t}-\frac{4\pi Q^2}{t^2}+\frac{8\pi\alpha^2Q^4}{5t^6}+
\frac{1}{3}\Lambda t^2-1 > 0.
\end{equation}
This inequality is satisfied for a wide range of values of the
parameters $m_0, Q, \Lambda,$ and $\alpha$. Here the additional term
diverges more rapidly than the Maxwell term and has the correct sign.
This spacetime behaves
in a similar fashion to the case with $Q=0$.
  
Finally consider the Born-Infeld Lagrangian. The constraint is
\begin{equation}
\frac{2m_0}{t}+\frac{1}{3}\Lambda t^2-1+\frac{8\pi}{a^2t}\int_0^t
\left[\sqrt{a^2Q^2+x^4}-x^2\right]dx >0.
\end{equation}
Since the integral is greater than zero for all $t> 0$
the inequality will certainly be satisfied if $m_0>\frac{1}{3}\Lambda^{-1/2}$.
In Born-Infeld theory the electric contribution remains finite and
does not present a problem as $t$ goes to zero.
For $m_0 >0$ this spacetime has similar properties to the case with
$Q=0$. It is possible to take $m_0=0$. For small $t$
\begin{equation}
\frac{2m(t)}{t}-1\simeq \frac{1}{3}\left[\Lambda-\frac{8\pi}{a^2}\right] t^2-1
+\frac{8\pi}{a}|Q| .
\end{equation}
Thus, we require that $|Q|\geq a/8\pi$. Now
let $f(t)=t(2m(t)/t-1)$. The derivative of $f(t)$ is given by
\begin{equation}
f^{'}(t)=\left[\Lambda -\frac{8\pi}{a^2}\right] t^2-1+\frac{8\pi}{a^2}
\sqrt{a^2Q^2+t^4} .
\end{equation}
If $\Lambda\geq 8\pi/a^2$ and $|Q|\geq a/8\pi$ then $f^{'}(t)>0$ for
$t>0$ and $2m(t)/t-1>0$ for $t>0$.
   
Equation (\ref{3}) determines the spacetime metric given $H(P^2)$. The 
reverse process is also possible. For a metric of the form
\begin{equation}
ds^2=-\frac{dt^2}{a(t)^2}+a(t)^2dx^2+t^2d\Omega^2
\end{equation}
the Hamiltonian is given by
\begin{equation}
H\left[\frac{Q^2}{t^4}\right]=\frac{1}{4\pi t^2}\left[\frac{d}{dt}(ta^2)
-\Lambda t^2+1\right].
\end{equation}
To be physically reasonable $H$ must reduce to the Maxwell
Hamiltonian in the weak field limit. 
\section*{Conclusion}
Exact cosmological solutions to the Einstein field equations coupled to
nonlinear electrodynamics, including Born-Infeld theory, were constructed.
These solution were produced by considering spherically symmetric solutions in 
which the roles of $r$ and $t$ are reversed. These spacetimes have an
initial singularity and constant time hypersurfaces have topology
$R\times S^2$. The radius of the two sphere is $t$ and the scale factor
of the other dimension diverges in some cases as $t$ goes to zero and
vanishes (or goes to a constant) in other cases.
At late times local observers would see the universe to be homogeneous
and isotropic.
Such solutions can be constructed
from the Schwarzschild solution with a positive cosmological constant and from
Born-Infeld theory. Maxwell theory does not lead to a cosmological
solution because the roles of $r$ and $t$ cannot be reversed for all
$r>0$.

\end{document}